\documentclass{kluwer}
\usepackage{graphicx}
\begin{document}
\begin{article}

\begin{opening}

\title{A New Challenge: Bar Formation and Secular Evolution in Lenticular Galaxies}
\author{Dimitri Alexei \surname{Gadotti}\email{dimitri@astro.iag.usp.br}}
\author{Ronaldo Eust\'aquio \surname{de Souza}\email{ronaldo@astro.iag.usp.br}}
\runningauthor{D. A. Gadotti \& R. E. de Souza}
\runningtitle{Bars in Lenticular Galaxies}
\institute{Departamento de Astronomia, Instituto de Astronomia, Geof\'{\i}sica e Ci\^encias
Atmosf\'ericas, Universidade de S\~ao Paulo --- Rua do Mat\~ao, 1226 -- Cid. Univers.
CEP 05508-900, S\~ao Paulo -- SP, Brasil}

\begin{abstract}
To further enhance our understanding on the formation and evolution of bars in lenticular (S0)
galaxies, we are undertaking a detailed photometric and spectroscopic study on a sample of
22 objects. Here we report the results of a 2D structural analysis on two barred face--on
S0's, which indicate that presently these galaxies do not possess disks. We discuss two possibilities
to explain these surprising results, namely strong secular evolution and bar formation without
disks.
\end{abstract}
\keywords{galaxies: evolution, galaxies: structure, methods: N-body simulations}

\end{opening}

\section{Introduction: do we really know how bars form?}

At the very beginning of galactic astronomy, barred galaxies were considered as anomalies in the
realm of the nebulae. Later, N-body numerical calculations (e.g., \opencite{hoh71}) as well as
analytic studies (e.g., \opencite{kal72}) gave us an idea on how bars are formed in galaxies,
namely via a global instability in dynamically cool disks. Then, the problem was how we could
explain why not {\em all} galaxies have bars, until a solution was found based on increasing the
stellar velocity dispersion in the disk, and/or adding a halo (see, e.g., \opencite{ost73}).

After a period of calmness, some recent studies brought up issues that have put us back again almost
at the starting point! \inlinecite{too81} already argued that a high central density disk will not form
a bar, which was confirmed numerically by \inlinecite{sel99}. Nonetheless, we observe barred
galaxies with dense centres (see \opencite{sel00}). Thus, how galaxies form bars remains an open
question. The answer is not only interesting for academic reasons, for it can give us important
clues and constraints to study the formation and evolution of galaxies (see, e.g., \opencite{gad01}).

The problem gets worse if we consider S0 galaxies, since these are not dynamically cool systems.
To tackle this problem, we have collected optical (B, V, R, I) and near-infrared (Ks) images of a
sample of 22 galaxies. Furthermore, we have also taken spectra along the major and minor axes
of the bar with a S/N high enough to obtain line of sight velocity distribution (LOSVD) profiles.
We report here partial results on two galaxies which have led us to surprising conclusions.

\section{Barred galaxies without disks?}

We have applied a 2D structural analysis code developed by \inlinecite{des97} to the optical
and near-infrared images of the SB0 face--on galaxies NGC 4608 and NGC 5701. The code
performs a bulge/disk decomposition and has been tested and applied in a hundred galaxies
(see \opencite{des02}). The results of this analysis indicate that these galaxies do not have
relevant disks; the major component is indeed the bulge (see Fig. 1).
Then, one is left with two possibilities.
Either the disks in these galaxies were almost completely destroyed by strong secular evolution
processes, or these galaxies have never had disks, which requires a new mechanism for bar
formation other than the disk instability.

To test the occurrence of secular evolution in these galaxies we may ask whether the stars we
see today in the bar once belonged to a disk. Thus, we have determined what would have been
a bulge/disk ratio for these galaxies before the evolutionary processes took place. This was done
assuming that all luminosity in the bar may be attributed to the pre-existent disk. We have found a
value for this ratio of $\approx 2$ which is consistent with the values found for normal S0's (see
\opencite{bin98}).

However, there is still the possibility that the disk has never existed. While bar formation without a
disk seems to be quite unusual, so it is an evolutionary process strong enough to destroy a disk.
Thus we have tested a very simple and powerful idea of a new mechanism for bar formation that
does not require a disk, but a non-spherical dark matter halo. With the {\sc nemo} package
(\opencite{teu95}) we have made N-body simulations of a spheroid embedded within the
potential of a rigid triaxial halo. Using parameters for the halo consistent with those found
in the literature, we show that for sufficiently elongated halos the spheroid may turn to a
structure which resembles very much the bars seen in the galaxies discussed here (see Fig. 2).

\begin{figure}
\tabcapfont
\centerline{%
\begin{tabular}{c@{\hspace{1pc}}c}
\includegraphics[width=2in]{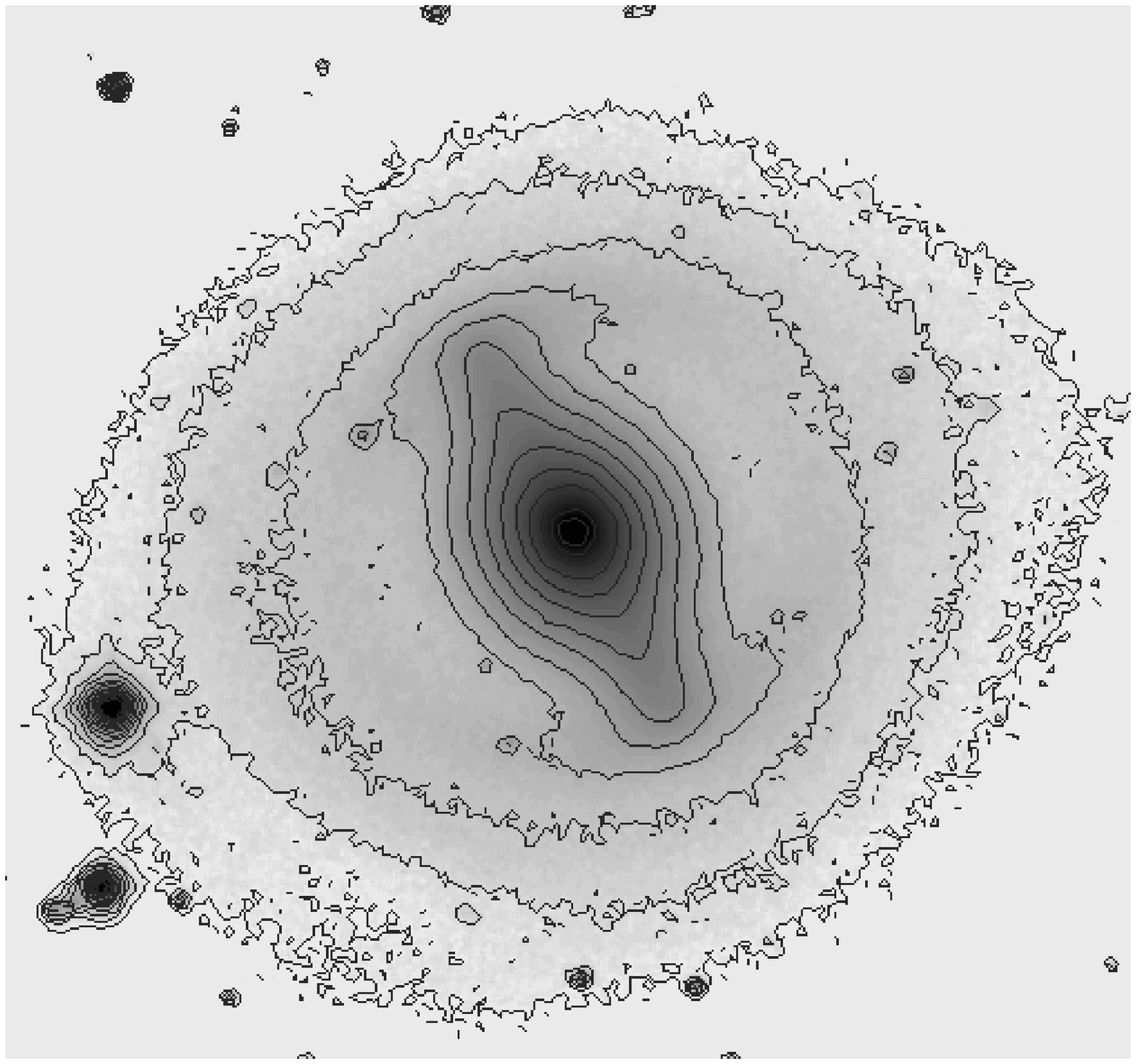} &
\includegraphics[width=2in]{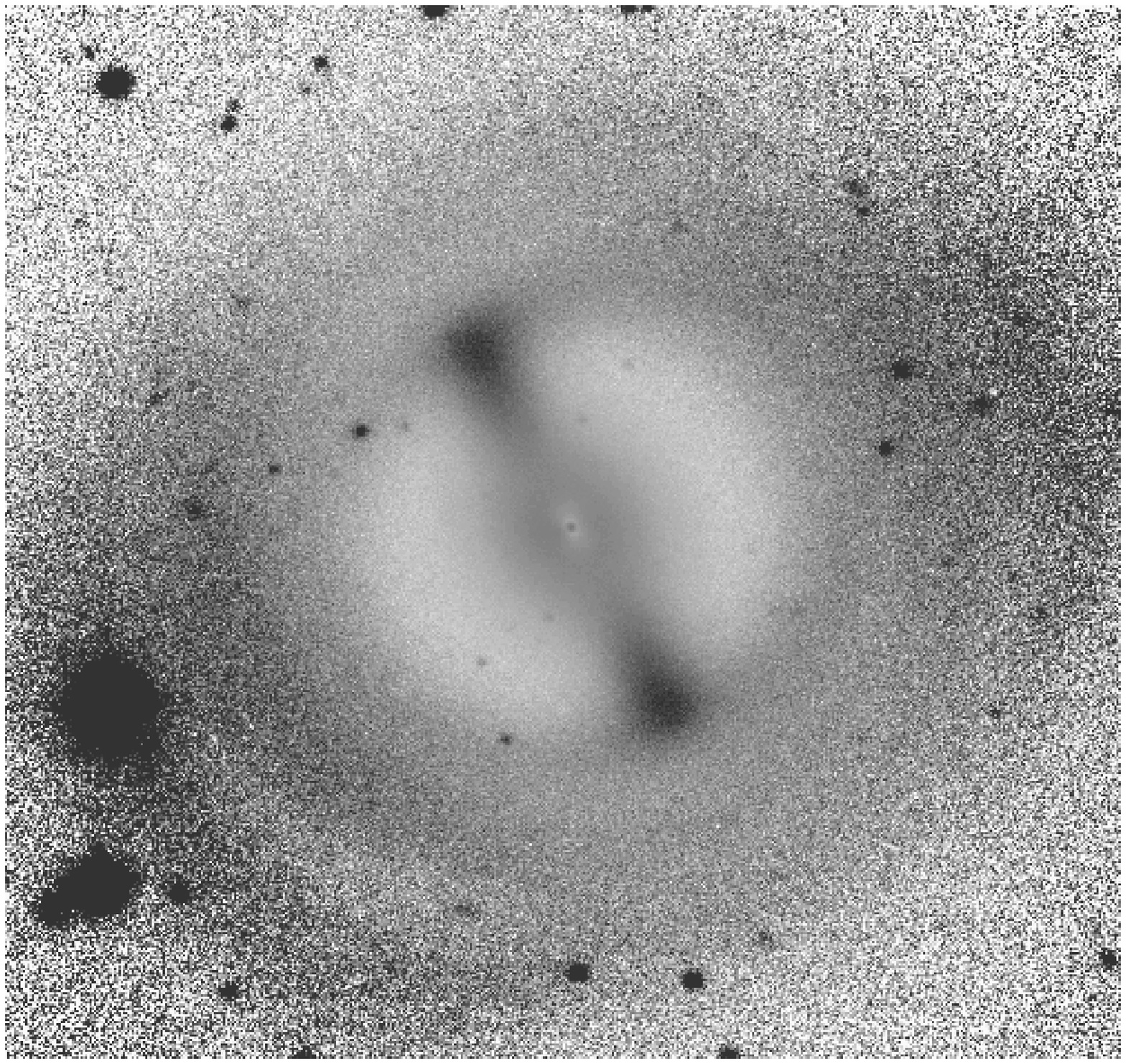} \\
\end{tabular}}
\vspace{1pc}
\centerline{%
\includegraphics[width=4in]{dgadotti.fig1c.eps}}
\vspace{1pc}
\caption{Results for the structural analysis on NGC 4608 in the V band. Similar results were
obtained for all the other optical and near-infrared bands, and also for NGC 5701. The top left
image is a direct image, whereas the top right image is a residual image after the subtraction
of only a bulge model. The images are 6 arcminutes each side.
Lower panels show radial profiles of relevant physical parameters:
surface brightness, position angle, ellipticity and the Fourier b4 component, after ellipse
fitting on the direct image (points with error bars) and on the image of the bulge model
(solid line). In the residual image it is easy to identify the empty region around the bar where
there should be a disk. Moreover, one can also identify an outer structure around the whole
galaxy which may be the outer remains of a pre-existent disk. In the radial profiles, all the
differences between the direct image and the model are caused by the presence of the bar
and the outer disk/lens. The radius $a$ is in pixels (100 pixels = 30 arcseconds).}
\end{figure}

\begin{figure}
\tabcapfont
\centerline{%
\begin{tabular}{c@{\hspace{0.5pc}}cc}
\includegraphics[width=1.5in]{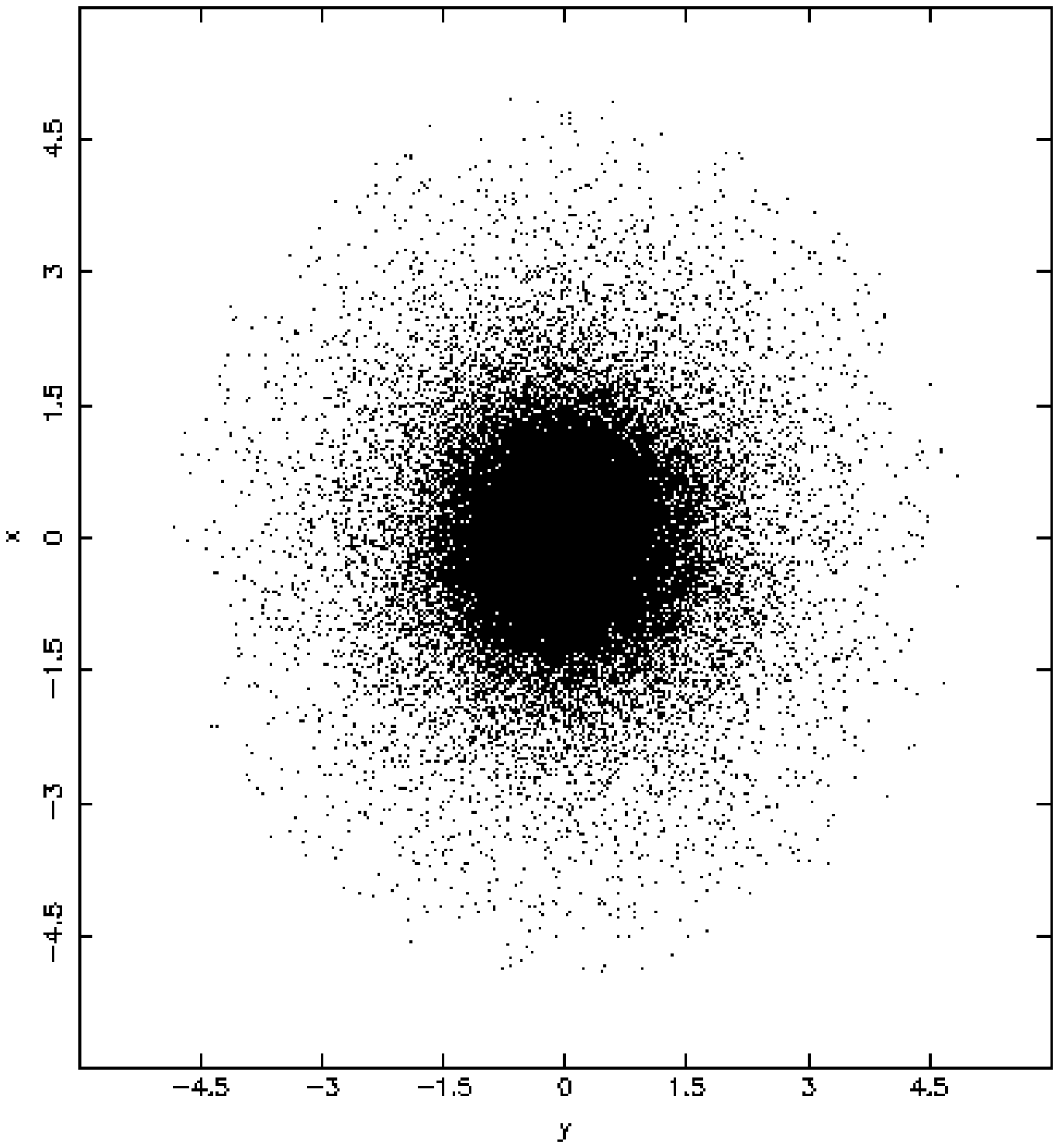} &
\includegraphics[width=1.5in]{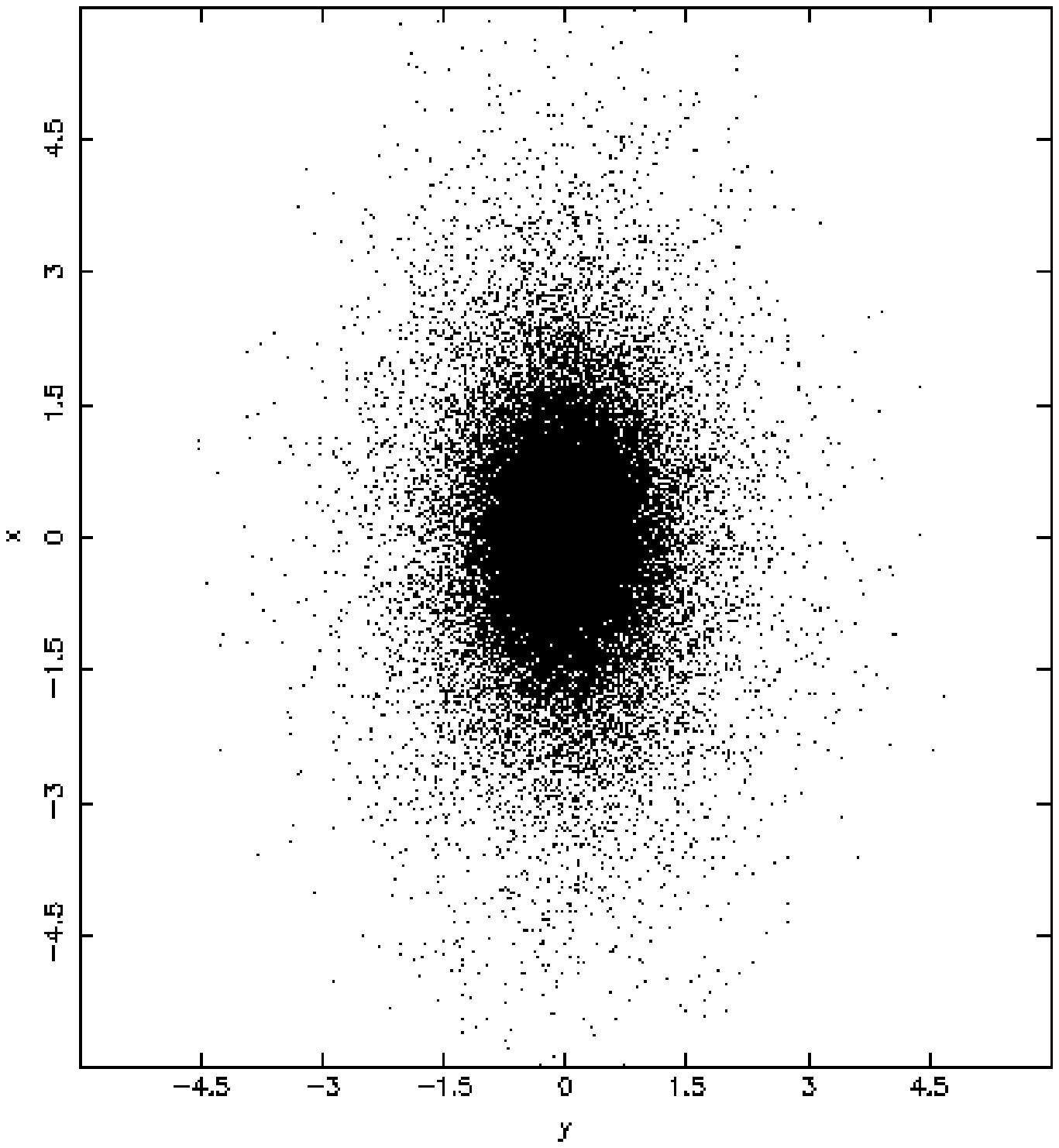} &
\includegraphics[width=1.5in]{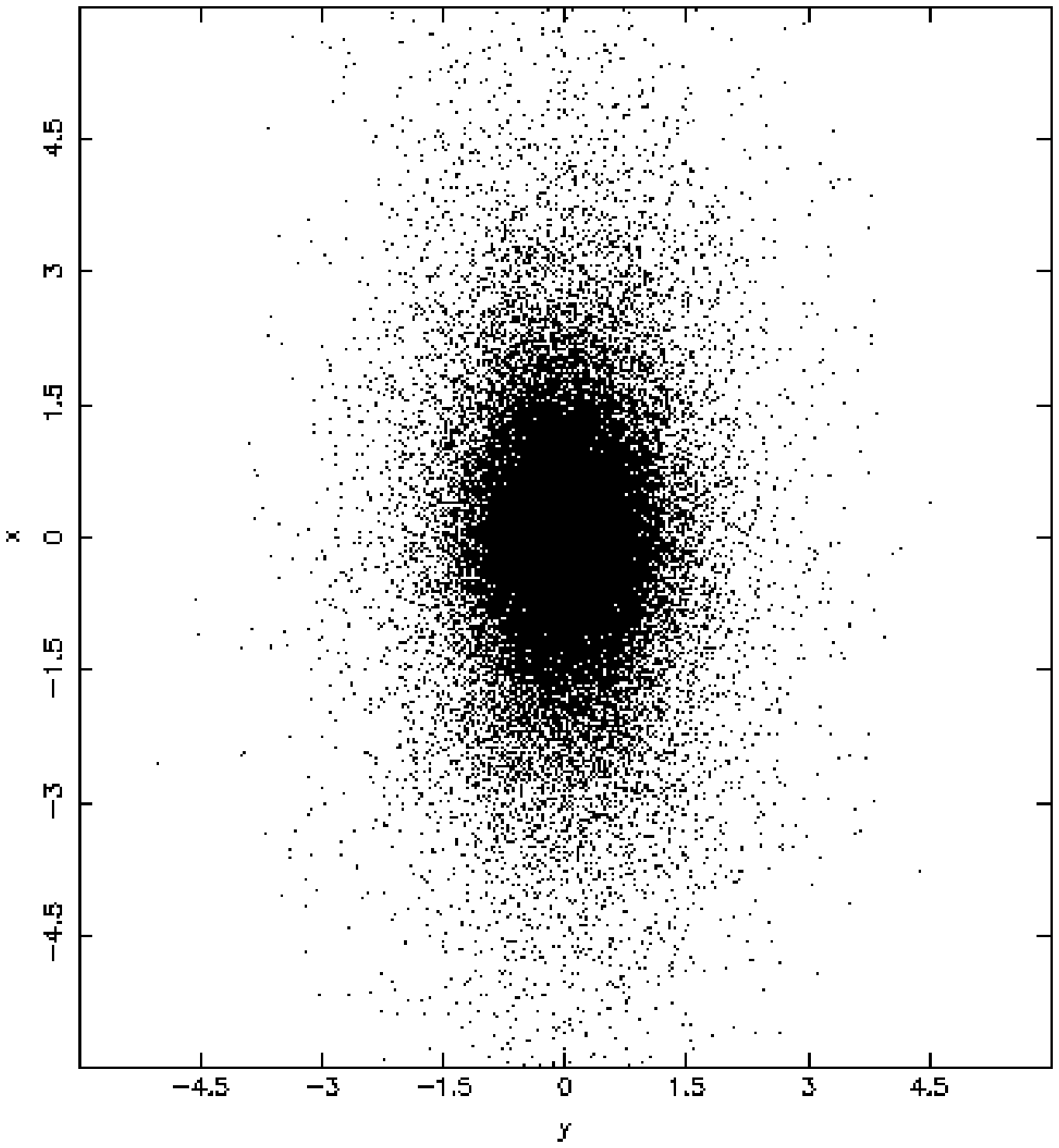}
\end{tabular}}
\caption{A Plummer sphere with $10^5$ particles (left) evolves for 1 Gyr within a triaxial
dark matter halo with axial ratio equals 2 (middle) and 3 (right). Compare the structures formed
with the bulge/bar in NGC 4608 (Fig. 1). The length unit is $\approx$ 3 kpc.}
\end{figure}

It is clear that the two possibilities are not necessarily mutually exclusive. This new mechanism
for bar formation may be responsible for the bars we see in S0's and may solve some of the
drawbacks of the disk instability to form bars in spirals. On the other hand, if secular evolution
is the answer to explain the absence of a disk in NGC 4608 and NGC 5701, then these are good
examples of how strong these processes may be and how seriously they should be considered.
Further details and results of this study may be found in \inlinecite{gad02}.

\acknowledgements
DAG would like to thank the organizers of this splendid meeting. Financial support from FAPESP
grant 99/07492-7 is acknowledged.

\end{article}
\end{document}